# Visions and Challenges in Managing and Preserving Data to Measure Quality of Life

Vero Estrada-Galinanes* and Katarzyna Wac*†
*Department of Computer Science DIKU
University of Copenhagen, Denmark
†QoL Technologies Lab, Center for Informatics,
University of Geneva, Switzerland

*Abstract*—Health-related data analysis plays an important role in self-knowledge, disease prevention, diagnosis, and quality of life assessment. With the advent of data-driven solutions, a myriad of apps and Internet of Things (IoT) devices (wearables, home-medical sensors, etc) facilitates data collection and provide cloud storage with a central administration. More recently, blockchain and other distributed ledgers became available as alternative storage options based on decentralised organisation systems. We bring attention to the *human data bleeding* problem and argue that neither centralised nor decentralised system organisations are a magic bullet for data-driven innovation if individual, community and societal values are ignored. The motivation for this position paper is to elaborate on strategies to protect privacy as well as to encourage data sharing and support open data without requiring a complex access protocol for researchers. Our main contribution is to outline the design of a self-regulated Open Health Archive (OHA) system with focus on quality of life (QoL) data.

## 1. Introduction

Health, quality of life and well-being are closely related. The World Health Organization (WHO) indicates that measuring quality of life can provide valuable information in medical practice, for improving the doctor-patient relationship, as well as for assessing the effectiveness and merits of treatments, in health service evaluation, in research and in policy making [1]. Quality of life (QoL) is multidimensional [2]. The qualitative and quantitative data collection of biological, physiological, psychological and social-environmental factors may tell the beholder unprecedented information about the data subject. Forging meaning from these factors in order to get a better quality of life (QoL) is undoubtedly a good reason to gather data. Concerns appear when we start discussing who controls or who processes these data collections. This dilemma raises privacy and security tensions [3]. At a larger scale, big data and the Internet of Things (IoT) open the opportunity to bring global health innovations. Though, researchers rarely gain access to these datasets that are kept behind commercial, ethical and legal barriers, and individuals may give away privacy based on illusions.

The use of non-medical devices to maintain or restore health [4], [5], reinforces the importance of patient-driven health care models [6] and consumer health care models [7] that include mobile health (mHealth) applications to track lifestyle habits, wrist-band trackers, or other kind of IoT/wearable devices. There is plethora of scientific peer-review pervasive computing systems for health care [8], [9], [10]. In addition, the International Data Corporation (IDC) predicted that the wearable market will continue to grow. It will have a compound annual growth rate of 18.4% over the next three years, i.e., 222.3 million shipments in 2021 [11]. Based on IDC's forecast, it is reasonable to expect a growing research interest in wearable devices applications for healthcare. With IDC collaboration, EMC projected that by 2020 healthcare data will require 2,314 exabytes of storage space if it continues with a growing annual rate of 48% [12].

We attempt to address *what strategies can be put in place to manage and preserve the sheer amount of personal health information* We strive for answers that create individual, community and societal values, encourage individuals to be more engaged in healthy behaviours without losing control over their digital traces, and empower researchers and citizen scientists with an ever-growing health data source.

Undoubtedly, access to data is essential to conduct data-driven research [13]. But academic labs often lack resources to gather data on a large geographical scale. A remedy could be to conduct research in association with an industry partner to gain access to big data repositories. But that introduces the risks of statistical bias since data collected by users of a single product brand, e.g. Apple smartwatch or Fitbit trackers, may not represent the general population. Another problem is that research projects that involve personal health data require ethical approvals. From our own experience, and that of others, the application process can be complex, time consuming and not necessarily successful.

The QoL Technologies Lab aims at improving the quality of life of individuals throughout their lives. The lab mission is to develop and evaluate emerging mobile technologies with the goal of assessing individuals' life quality as it unfolds naturally over time and in context, and improving it at all stages of life. Since a sedentary life is a major cause of chronic diseases [14], we collect data from many sources for understanding better the health implications of lifestyles behaviours. We aim to develop algorithms that learn patterns from patient-generated data to increase personal well-being, activity, promote healthy ageing and improve the conditions of ambient assisted living. To sustain innovation we need data to conduct measurements covering physical health, psychological, social relationships and the environment.

There is general agreement that QoL and well-being measurements have to be sought from the individual own's perception of quality of life and health. In a sense, life-logging technology and self-quantification is at the core of patient-centric research. More traditional clinical studies use clinician-reported (ClinRO), performance-based (PerfO) and self-reported (SRO) outcomes [15]. Professionals assess activities in PerfO, while the participant is required to fill SRO. The introduction of wearable devices in the healthcare sector is under discussion by several groups. A main argument is that patient-supplied data open the opportunity to develop new decision-making metrics. But unlicensed mHealth applications and wearable devices need some form of attestation to verify their effectiveness. Meanwhile, QoL researchers like us leverage technologies such as wearable devices and smartphones for taking measurements in daily life activities. Researchers collect data with the participant's consent using ad-hoc applications or APIs that connect to cloud-based repositories.

There are several drawbacks in the above described model. PerfO are momentary and expensive. SRO may be inaccurate due to perception bias. It is difficult to visualise the multidimensional aspects of quality of life with mHealth apps if they focus on narrow aspects. Sometimes individual efforts to collect data are lost, especially when the subject does not have the control on his/her data.

From the perspective of the study participant, data is usually kept in silos. Universities, institutions, and research labs like us may have policies to retain data but those archives are not meant to be a backup for the participant. For instance, if the participant wishes to access her/his data in future, it will require to follow complex procedures. Similar complications appear with patient- and consumer-generated data from wearables and other sensors used for life tracking. The common practice is that devices synchronise their data to cloud-based storage systems directly or via smartphone applications. It becomes difficult to keep track all together the data from multiple sensors all along the life span of the person. Our conjecture is that the lack of a reliable storage platform to keep personal health and health-related records may have a repercussion in the user engagement with mHealth technologies and health self-management.

From the perspective of QoL researchers, the access to data is limited. Clinical studies usually involved large cohorts (in the order of thousands) and some studies extend across 10-30 years. Conversely, QoL studies extend on average 6 months and usually involved 20-30 participants. To the best of our knowledge, there are no such large company-independent data banks for data gathered with wearable devices, i.e., wearable data. Large biobanks have gather wearable data from large cohorts but only for short periods, e.g., UK Biobank includes a repository of 1-week wearable data used by more than 100k participants. Gathering data for larger periods and enable flexible mechanisms to voluntary share data may speed up innovation. To sum up, we identify two key aspects for bringing data-driven innovation in the QoL domain: 1) increasing the user engagement to quantify and track QoL measurements; 2) facilitating data gathering and data sharing.

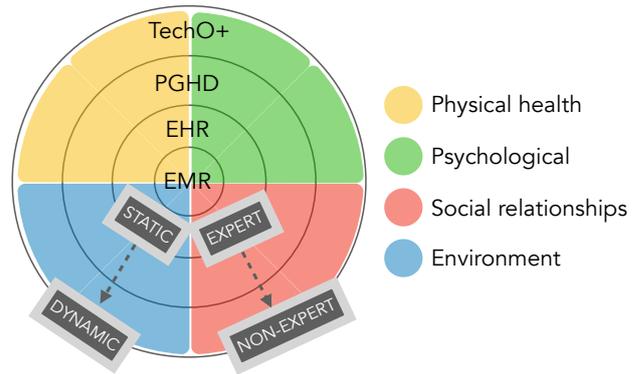

Figure 1. The wheel of personal health information.

The main contribution of this paper is to outline the requirements and design choices for an open health archive (OHA) as the means for keeping quality of life indicators in the hands of the data subjects. We share our overall perspectives and experience on the challenges of gathering and sharing quality of life (QoL) data. Finally, we discuss what it takes to create an archive that enables prospective and retrospective medical studies based on non-clinical data collected from birth.

## 2. Personal Health Information

ClinRO along with PerfO and SRO are key information sources for healthcare. By personal health information, we go beyond those key sources to include data generated by individuals with different levels of health expertise and expand the conventional health data model with quality of life data models. Leveraging this information can help QoL researchers to get a holistic view of the person's quality of life changing over time. In this section, we review the different elements of personal health information and elaborate on the privacy, security and ethical challenges to get more value.

### 2.1. Expert Data and Dynamic Data

Figure 1 highlights the characteristics of what we call the wheel of personal health information. Its inner sections correspond to records generated by experts. Its outer sections incorporates data generated by non-experts. We assume that the integration of non-expert sources contributes to reduce data acquisition costs and increases the possibilities to capture with higher fidelity the dynamic nature of quality of life [16]. Hence, the inner-outer wheel bridges the gap between static and dynamic measurements.

The time interval between measurements taken by experts can be long (with the exception of patients with critical health conditions who are closely monitored in hospitals). The use of dynamic data is encouraging to understand if interventions are working or not. In other words, laying data out over time is useful to understand trends and improve analytics. Thus, measurements taken at home may unveil relevant patterns even if the measuring device was not designed for medical purposes. Furthermore, in the domain of quality of life, Evans and a number of other researchers have argued that objective measurements should rely more on the actual manifested behaviour and current circumstances of the person compared with an external baseline [17]. Overall, dynamic data can provide a more comprehensive quantification than static data or self-reports used alone.

The wheel follows the distinction between "medical records" and "health records" given by the Office of the National Coordinator for Health Information Technology (ONC), U.S. Department of Health and Human Services (HHS)[1]. The innermost layer are electronic medical records (EMRs), a digital version of the traditional paper patient's history. Electronic health records (EHRs) expand the value of EMRs. They contain information from all the clinicians involved in the patient's care. Patient-generated health data (PGHD) expand the value of EMRs even more with health-related data created, recorded, or gathered by or from patients, family members or other caregivers. They include but are not restricted to patient-reported outcome measures (PROMs), observer-reported outcomes (ObsRO) and proxy reported outcomes (ProxRO). Examples are self-reported questionnaires, face to face interviews and telephone interviews about physical, mental, and social health. Recently, a relationship model between patient-generated data and other source data was proposed [15]. In that work, technology-reported outcome (TechO) is vaguely mentioned as part of the clinician-reported outcomes (ClinRO) branch and directly linked with performance-reported outcomes (PerfO), which are assessments done by clinicians in close cooperation with the patient. However, the work presented in here proposes TechO+, to distinguish and take into consideration the data that is measured by sensors, trackers and other IoT devices either worn by the user, or available in the user's environment, e.g., smart home or a smart car, specifically used by the individuals interested in quantifying different aspects of their life. Thus, to leverage these records, TechO+ is included in the wheel as the outermost layer.

## 2.2. Privacy, Security and Ethical Challenges

Trust is a key condition to fulfill in the governance of health data in order to generate innovations in digital health [18]. Common solutions to gather personal health information are centralised, meaning that the data subject is at the mercy of technology providers, researchers, healthcare providers, doctors, wellness centers, etc. All of them have in place their own systems to manage and store data from their patients/consumers. Systems must comply with law but there is still room for arbitrarily decisions in privacy and security policies that could affect the data subject. In addition, the reading level of privacy statements is intelligible only to a minority [19]. As a result, the data subject is often powerless.

Individuals may give consent to gather data based on the expectations to address health concerns. In this context, businesses keep exploiting opportunities although it may affect individuals [3]. As stated in the same paper, in an algorithmic decision-making world, the individuals are exposed to privacy and security tensions. In the context of healthcare, an example could be that a patient's privacy is violated with an excessive monitoring justified by an algorithm that categorises certain individuals as more likely to commit suicide.

Thus, a significant barrier for exploiting personal health information in the benefit of society is trust, where much is requested from individuals. Conversely, there are multiple examples of negligences and/or abuses of trust that can lead to vulnerabilities and/or security breaches to gain more information about a subject [20], [21], [22]. If human blood is for traditional (analogue) healthcare the equivalent to human data for the digital healthcare, we can talk about *human data bleeding*. The human body is a source of big data, especially in the context of health. For example, a single X-ray accounts for 30MB of data, a mammogram for 120MB, 3D MRI for 150 MB, while 3D CT scan for 1GB. Overall, IBM estimates that there are 150 exabytes of healthcare data in US alone (electronic patient records, compliance, insurance, patient care data) and medical image archives are increasing 20-40% annually [23]. An extreme case of continuous-high resolution health monitoring is given by University of Ontario and IMB who have developed a service for monitoring of premature-born babies, known as preemies, with 16 different data channels for heart rate, temperature, respiration rate, oxygen saturation, blood pressure, which accounted to 1260 data points per second (accounting to 40 kbps assuming 4Bytes/data point) [24].

Regulations such the General Data Protection Regulation (GDPR) - (EU) 2016/679 to protect individuals within the European Union and the European Economic Area are positive steps. GDPR imposes obligations to data controllers and data processors, among other protections, to defend the individual's right to access, move and forget data. However, there is no much experience with these new regulation neither with its implementation. As stated by the Open Data Institute (ODI) in its guidance [25]: "Data-related activity can be unethical but still lawful." After all, we wish to avoid human data bleeding and, accordingly, not only conform with *controlled bleeding of human data*.

GDPR was designed with centralised architectures in mind, therefore, implementing GDPR can be quite complex if data is distributed in many systems. It is necessary to evaluate all source of personal identifiable information, e.g. legacy systems, backups, etc. To remediate and facilitate im-plementation, an apparently reasonable advice is to eliminate unnecessary sources of personal identifiable information and centralise even more the solutions. Cloud companies might find easier to implement GDPR than traditional companies [26]. But such approach may lead into even more massive data collections administered by few companies.

---

1. https://www.healthit.gov/

Centralised solutions, administered and regulated by third-parties, are not the only possible architecture for storing personal health information. Though, in the healthcare domain other alternatives had been rarely explored until blockchain became a popular alternative. In fact, there are many different solutions that can give more control to the data subject. Data could be stored locally (at home) in equipment owned and administered by the data subject. It's worth to mention that storing data outside central silos is not an impediment to get valuable information from these resources. Researchers showed that it is possible to apply Bayesian predictive models [27] and deep learning methods [28] in a more collaborative and distributed environments to reduce privacy concerns.

Completely decentralised solutions based on blockchain or other distributed ledgers to store health data are trying to find the way in the complex regulated healthcare space. When storing data in blockchain-based storage systems the first question to ask is whether is a permissionless (public) or permissioned (private/consortium) blockchain. Permisionless blockchains distribute data across all nodes in the system. Nodes contribute with resources, e.g., storage, bandwidth, computation and electricity, (often in exchange of a fee). But keeping personal data in a public blockchain conflicts with GDPR privacy requirements. Encrypted data is considered under the umbrella of GDPR, since the data subject can be indirectly identified [29]. In permissioned blockchains, an administrator can control and grant permissions. These are still alternatives to traditional databases, but the data subject may become powerless again due to the centralised administration. The combination of blockchain and off-chain storage is becoming a common answer to address privacy concerns [30]. Personal data can be stored in a separate location (where data can be modified and deleted) and linked to a blockchain with a pointer. Given the many implementations, this topic deserves careful consideration and further research.

### 2.3. Individual, Community and Societal Values

Health data is being jeopardised by patient portals and mHealth apps that populate the market but disappear in the short time. Common pitfalls are lack of simple procedures to move/share data, lack of transparency, data is difficult to navigate. Databases that are fragmented and hidden between walls difficult coordination among all healthcare actors and discourage patients and health consumers. As a result, individuals may become less engaged in healing or improving his/her own health and wellbeing. But modernising this area with improved data access methods requires to consider: a) technological, b) financial and budgetary, c) legal and policy, d) institutional and managerial, and, e) cultural and behavioural factors [13].

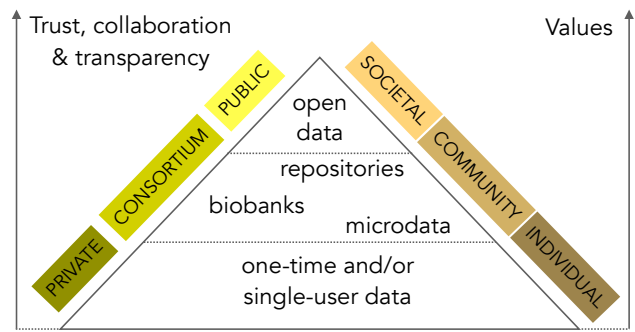

Figure 2. The hierarchical value of health data.

As first step toward the design of a solution that can provide long-term value, we assign a hierarchical value to health data. Based on this hierarchy, systems that manage and preserve health data can be classified in a pyramid of three non-exclusive categories, as illustrated in Figure 2. *Individual value* is attainable by systems designed for one-time use, e.g., a particular clinical study conducted with the help of a mHealth app, or for single-user, e.g. commercial applications that satisfy the needs of individuals. *Community value* is possible when data is kept in well-curated institutional repositories, biobanks or public microdata collections. Temporal, economical, geographical and other restrictions may apply, consequently, benefits are subscribed to a group of individuals. *Societal value* is viable with open, and ideally raw, data. Systems that gain space in the upper part of this pyramid can bring social innovation since quality of life measurements can have a significant impact in many fields that need creative solutions to meet social goals. Some of these fields are the rising incidence of long-term conditions, behavioural problems, happiness, rising life expectancy, and more examples given in the literature [31].

Certainly, the need for more trust, collaboration and transparency is a requirement to attain community and societal value. For this to happen, systems that manage and preserve health data offer different levels of requirements. Even if open data is our ultimate goal, privacy at the pyramid's foundational level should be preserved too.

### 2.4. Existing Solutions

Table 1 presents representative non-profit community projects and commercial solutions to manage and archive personal health data. Open Humans is close to our vision and values but data resides in US commercial servers. That may raise questions regarding privacy and ethics regulations for EU residents and other countries. Dat protocol does not target the end-user and none of the dat-based existent applications developed by the community specialised in the core QoL domains. Commercial solutions aim at delivering product/services to individual and communities who can pay for their solutions. With the exception of the Dat protocol

TABLE 1. EXISTENT COMMUNITY/COMMERCIAL SOLUTIONS.

| SOLUTION | TYPE | DESCRIPTION |
|---|---|---|
| Open Humans | non-profit community project | The Open Humans project aims at providing a path to share data, such as genetic, activity, or social media data, with researchers. Data is ingested using 19 connectors to different external sources. Users can store data privately to archive and analyse data or opt to contribute data to projects proposed by the community. The website and other tools are open source and the storage service is free. Datasets are stored on Amazon's US Web Services. |
| Dat protocol | non-profit community project | The Dat protocol is a data sharing protocol supported by a decentralised architecture. Its initial goal was to archive and share scientific data, although the project is evolving and expanding its domain. |
| Microsoft Health Vault | commercial | It is a cloud service to store health data oriented to businesses (including hospitals, pharmacies and lab testing companies) and consumers. Consumers can use the service to gather, store, use and share data of themselves, children and other family members. |
| Wolfram Data Drop | commercial | Data Drop is a commercial solution to upload data of any type to Wolfram's cloud-based repository (databin) for visualisation, analysis and queries. It ingests data using a large collection of connectors. Users can administer the databin by assigning its creator, owner, and one or more administrators and grant public/private read/write permissions. |
| Umbra Lifeography | commercial | Lifeography is a commercial solution specifically to keep patient controlled health records (test results, procedures, medications, notes, lifestyle stats. It requires partnerships with doctors, clinics, hospitals, and wellness facilities. It includes support to parents to collect baby's measurements prior to delivery, through birth and as the child grows. |

that makes possible to store data in a decentralised way, all the others use cloud-based storage.

In recent years, research on cooperative health systems to change the way health data is managed and shared has attracted attention. Previous work [32] proposed a cooperative health system that is closed to OHA spirit, i.e., empowering the data subject with a platform to store, manage, and share data supported by a non-profit organisation. They promoted the idea of cooperatives at the local level and a federation of cooperatives at a global level. Pilot studies based on a cooperative model have been built in different places. In the late summer of 2010, an organization tried to build a health bank record for a community in Phoenix, Arizona (USA). Their aim was to engage 200k users (5% of the population), but due to insufficient registrations the project was abandoned in April 2011. Consequently, researchers explored the challenges of creating cooperative health systems [33], [34]. One of the requirements for those cooperative health systems was a one time fee and many early adopters. But the experience in Arizona revels that not many people (only 20%) may be willing to pay for it. In Switzerland, the MIDATA.coop initiative decided for non-for-profit organizations and since 2015 they are pushing for democratization of personal data [35] with different national cooperatives. In Netherlands, the personal health train initiative promotes the use of FAIR principles (we will discuss them later in this paper) and citizen-controlled health data lockers [36]. At the time of writing this paper, the authors did not disclose publicly any detailed information of the health data lockers. Based on the scarce information available, it is not very clear if the citizen will keep control, e.g., have a copy of the data. On the contrary, data is shared among partners to integrate datasets without leaving the source.

## 3. Open Health Archive (OHA)

This section outlines the requirements and design choices for the Open Health Archive (OHA) system. The main concept in this design is to preserve the health-related data generated throughout the life of an individual without giving away data ownership while promoting open data and data sharing. We argue that in order to have control over personal data collections, the person should participate in the system administration and ideally he/she should have physical access to the storage devices. Encouraging data subjects to take control on their health data requires more engagement and more responsibility. But open-source community efforts can help to reduce the burden for end users.

Our vision for OHA is to put emphasis on the creation of value in a pyramid structure that benefits the individual, communities and society. Perhaps some persons donate data for research altruistically, but in general, we think that the individual needs have to be carefully considered to generate sustainable benefits to communities and societies. Otherwise, the engagement factor may be low, causing a decrease in the number of registrations and/or increasing participation attrition. We promote gradual openness to protect and respect the privacy according to individual requirements while making easier the path to share data whenever the data subject decides it (including posthumous donations). We discuss strategies that can create value such the creation of a knowledge base and pseudo-anonymous profiles for each of the core QoL domains.

A distinguishing factor for OHA is our position regarding data storage. Most of the existing solutions are based on cloud storage, but there are three main reasons why we think that cooperatives should consider other alternatives such as do-it-yourself (DIY) or hybrid storage solutions based on decentralised architectures. A hybrid solution combines components located in-house with other remote components. First, the economics of long-term archives are different from models develop to store data that has a lifespan shorter than the life of the storage device [37]. Cooperatives could afford the upfront investment with the registration fee. Second, the use of public cloud storage means that the provider has some control on the data. Instead, on-premise solutions can keep the data where it should be, in the hands of the cooperative and, more specifically, in the hands of the data subject only. Third, the public cloud is not error-free, outages in health data systems can be potentially life threating. Thus, such systems should provide high availability and reliability.

We envision a self-regulated system that shares computer resources in a decentralised storage system, assures patient/consumer-centric data management administration and provides for each user a monolithic view of their life-tracking log. Developers could offer services or applications running on top of OHA but, at least, OHA system's core should be an open-source project. In this paper, we put more emphasis in the two outermost layers of the wheel, PGHD and TechO+ data, because the data subject already has some control over these data collections, e.g., Fitbit backups. Accessing medical records and stored them under the data subject realm is a very political and complex topic, not for short-term research. However, the ultimate goal of OHA is to preserve the data generated at every layer of the wheel of personal health information.

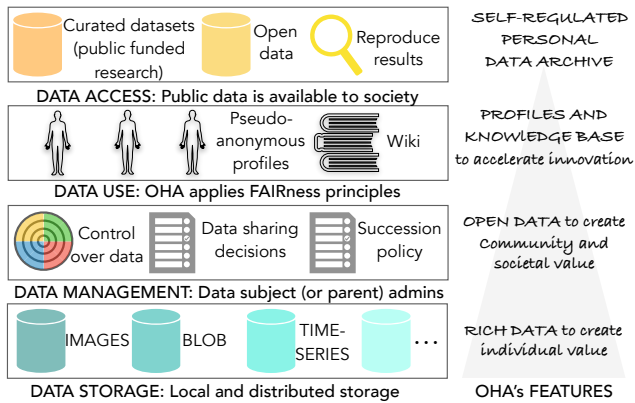

Figure 3. Overview of the Open Health Archive (OHA).

Promoting open data and data sharing does not mean to ignore that citizens need secure systems to store their personal health information, and that most of the data collections kept in OHA may remain private. The system should store data privately and offer policies that enables open data gradually. Individuals may opt to participate in research studies and donate data to communities. Based in the premise that data sharing can transform health research, OHA should encourage researchers to share results with the community and to reproduce studies made by other researchers.

Building a cooperative, self-organised health archive system is a complex and long-term endeavour. We believe that the initiative to create OHA must be taken by a non-profit organisation that acts in the best interest of its members and adopts conflict of interests policies to assure that OHA remains open and has a transparent development and administration. This section elaborates more about our vision and outlines what can be the initial OHA requirements.

### 3.1. Preliminary specifications

At this stage of the research, our specifications are necessarily speculative since OHA is not yet implemented. However, given the crescent interest on health data governance and stewardship, this subsection presents some of the novel aspects of our design. We propose and examine the following OHA's core features, an overview diagram is shown in Figure 3:

**3.1.1. Rich Data.** OHA moves the focus of attention from big data to rich data. Rich data is constructed by aggregate inputs from different sources (or connectors) to fill in the layers and quality of life subdomains that define the wheel of personal health information. These rich data repositories per individual give the possibility to design a monolith life-tracking log. Gathering data from different source is possible via the integration of open standards, e.g., open mHealth standards [38]. Thus, the individual can monitor themselves their quality of life, and provided that they have the right instruments designed by experts in the field, visualise the correlations between different factors that may impact in their health. Researchers and healthcare professionals that are granted total or partial access to the monolith log will be able to conduct retrospective or prospective studies that take into consideration more dimensions of the quality of life and its dynamics.

**3.1.2. Individual Values.** Having all data in the same archive platform makes possible comprehensive analysis. Time travel visualisations through lifestyle patterns will be possible with time series data. OHA enables learning from patients with similar health concerns and learning from individuals with similar lifestyle patterns. It may be possible to conduct performance tests, e.g., 6 minute walk test, and compare results with previous personal measurements, community members or standard references. Participants may expect to receive economical or other benefits from sharing data.

**3.1.3. Community and Society Values.** Researchers will have access to a large-scale repository built with donated data. Scientific findings based on the open repository can be reproduced by multiple scientists. OHA will stimulate investigations based on the open repository to accelerate health innovation

**3.1.4. Pseudo-Anonymous Profiles and Knowledge Base.** OHA repositories are a source to generate statistical knowledge using validated instruments such WHOQOL-BREF 36 self-questionnaire. Users may participate in surveys to contribute to a large-scale statistical knowledge base and get feedback without fears of hidden data sharing policies. Optionally, this information is included in a pseudo-anonymous vector profile that is used by OHA's knowledge discovery mechanisms to connect people with compatible needs, e.g. a patient with type II diabetes and a researcher studying the effects of depression on patients with diabetes type II.

OHA is a computational stakeholder powered by algorithms developed by the community to help humans with the management and archival of personal health information. Non-experts are guided to gather more reliable and reusable data for investigations. Mechanisms to discover public data are put in place. Towards that goal, data stewardship follows good practices for scientific data management such the FAIR Guiding Principles, a recently published recommendation to "support knowledge discovery and innovation" [39]. FAIRness data means that data and/or metadata is findable, accessible, interoperable, and reusable. Yet, data sharing is not mandatory in OHA. Data is stored privately by default.

OHA allows gradual openness to promote data sharing and open data. Users may opt to share partial information with formal and informal caregivers, specific community groups or with the whole society. Sharing can be done at any time (for example, information can be shared after the appearance of a disease), including automatically after subject death with previous consent. To protect open data, OHA adopts a small number of possible licenses to donate content. End users may opt among standards such Open Data Commons Public Domain Dedication and License (PDDL), Open Data Commons Attribution License (ODC-by), Public Domain Dedication (CCZero or CC0), or Community Data License Agreement (CDLA). In addition, succession policies are put in place so that users state clear instructions regarding data deletion or posthumous data donation to family members and/or to health research institutions. Decisions are modifiable at any time by the data subject. Succession process starts with a valid death certificate.

OHA is supported by a hybrid infrastructure that combines local storage own by the user and distributed storage shared or leased to the user. The use of traditional cloud storage service is discouraged to avoid single points of failure, expensive long-term storage costs and privacy threats often associated with central administrations.

The general-purpose wheel of personal health information is composed of subsystems that ingest data from different sources. Any new subsystem incorporated in OHA need to state clearly its specific data policies. A possible list of questions that should be clearly addressed are taken from a previous study [40]: 1) Who owns the data? 2) What type of data, and how much data, should be stored? 3) Where should the health data be stored? 4) Who can view a patient's medical record? 5) To whom should this information be disclosed to without the patient's consent?

OHA collects data that (optionally) spans the whole life of a human. Parents or guardians are named data administrators before the data subject achieves legal adult age. The suggested data retention period is 100 years.

## 3.2. QoL Domain: A Case for Managing Sleep Data

Among factors influencing the QoL of the individual, WHO mentions sleep and rest. As well known, sleep is essential for well-being, health, and performances. Humans need 7-8 hours of consolidated sleep, preferably overnight [41]. However, many individuals sleep one or two hours less and sleep pathologies affect in the USA alone about 70 million people, approaching epidemic levels. Long term sleep disturbances and pathologies correlate with higher risk of chronic illness like hypertension, diabetes or CVD risk later in life. Within OHA we assume sleep as one of the vital signs that the individual will collect data for; not just episodically but in a structured, longitudinal way. Table 2 presents the considered sleep-related datasets being collected and their approximate size.

## 4. Conclusion

Data subjects have been historically treated as second-class citizens of data management and archive solutions. Many enthusiasts life trackers collect data per years but the friction of preserving their collections remains. mHealth apps are usually conceive with short-term goals. Many solutions are backed in cloud storage. Arguably, the cloud is not the ideal place for personal health information. To address these concerns, we propose the creation of an open health archive (OHA), a collaborative platform for management and archival of personal health information, and present the design philosophy behind OHA.

TABLE 2. PARAMETERS CONSIDERED FOR THE SLEEP ASSESSMENT WITHIN THE QoL.

| DATA | HIGH-LEVEL | LOW-LEVEL DESCRIPTION | APPROX. SIZE |
|---|---|---|---|
| EMR/EHR | health context (ClinRO) | diseases: esp. sleep disorders (e.g., insomnia), medication usage, etc. | 1 MB/year |
| PGDH | sleep quan/qual (PROM + ObsRO): last month | Pittsburgh Sleep Quality Index (PSQI [42]): sleep latency, sleep duration, habitual sleep efficiency, sleep disturbances, use of sleep drugs, daytime dysfunction and subjective sleep quality | 0.5 MB/month |
| | influencing factors: usuall | work shift schedule, usuall sleeping context (if sleeping alone/not, with a pet, young children etc.), if smoking, etc. | 0.5 MB/month |
| | influencing factors: last night | pain or fatigue levels, sleep location (e.g., home, hotel), meditation, reading, TV, late caffeine intake, alcohol, sexual activity, etc. In the morning, an individual may be asked to assess his/her energy levels, fatigue, pain, dreaming, interruptions and overall perception of the night of sleep. It may be also beneficial to know if the wake up was natural or initiated by an alarm clock or other external stimuli (bed partner, pets, children), etc. | 20B/night |
| TechO | sleep data collected within the sleep laboratory (leveraged for diagnostic): one night | Polysomnography: the brain (EEG, 256Hz), eyes (EOG, 8Hz), muscles activity and movement (EMG, 128 Hz), and heart (ECG, 256Hz) activity, and the individual oxygen saturation levels, body temperature and respiration rate. Additionaly to brain activity patterns, sleep latency, sleep architecture, (sleep stages and sleep cycles), sleep fragmentation, Wake Time After Sleep Onset, awakenings, Periodic Limb Movement Index, Respiration Disturbance Index, Snoring Index or circadian aspects | 144 MB/night |
| TechO+ | sleep data collected by individual, anywhere, long-term: every night | (i) smartphone: body movements (based on accelerometer; proxy for sleep/awake) based on build-in sensors (ii) smart watches (e.g., Nokia), bracelet- (e.g., Fitbit) or ring-type (e.g., OURA) devices: body movements - sleep duration, start/end time, sleep stages and cycles, awakenings, time in bed, minute-by-minute heart rate (and its variability), respiration rate, body temperature (iii) bed-side devices (e.g., S+) for sleep and environmental conditions assessment (ambient temperature, humidity, noise, light). | 5 MB/night |